\documentclass{aastex63}

\shorttitle{Experiments on Deuterated QCC}
\shortauthors{Mori et al.}
\begin{document}

\title{Laboratory measurements of stretching band strengths of deuterated Quenched Carbonaceous Composites (D-QCC)}

\correspondingauthor{Takashi Onaka}
\email{onaka@astron.s.u-tokyo.ac.jp}

\author[0000-0001-9985-5866]{Tamami Mori}
\altaffiliation{Present address: Recruit Holdings Co., Ltd., Tokyo 106-6640, Japan}
\affiliation{Department of Astronomy, Graduate School of Science, The University of Tokyo, 7-3-1 Hongo, Bunkyo-ku, Tokyo 113-0033, Japan\\
}

\author[0000-0002-8234-6747]{Takashi Onaka}
\affiliation{Department of Physics, Faculty of Science and Engineering, 
Meisei University, 2-1-1 Hodokubo, Hino, Tokyo 191-8506, Japan\\
}
\affiliation{Department of Astronomy, Graduate School of Science, The University of Tokyo, 7-3-1 Hongo, Bunkyo-ku, Tokyo 113-0033, Japan\\
}

\author[0000-0001-7641-5497]{Itsuki Sakon}
\affiliation{Department of Astronomy, Graduate School of Science, The University of Tokyo, 7-3-1 Hongo, Bunkyo-ku, Tokyo 113-0033, Japan\\
}

\author[0000-0002-5436-9845]{Mridusmita Buragohain}
\affiliation{Department of Astronomy, Graduate School of Science, The University of Tokyo, 7-3-1 Hongo, Bunkyo-ku, Tokyo 113-0033, Japan\\
}

\author[0000-0003-0124-208X]{Naoto Takahata}
\affiliation{Atmosphere and Ocean Research Institute, The University of Tokyo, 5-1-5, Kashiwanoha, Kashiwa, Chiba 277-8564, Japan\\}

\author[0000-0002-3305-5644]{Yuji Sano}
\affiliation{Atmosphere and Ocean Research Institute, The University of Tokyo, 5-1-5, Kashiwanoha, Kashiwa, Chiba 277-8564, Japan\\}
\affiliation{Center for Advanced Marine Core Research, Kochi University, B200 Monobe, Nankoku, Kochi 783-8502, Japan\\}

\author[0000-0001-6328-4512]{Amit Pathak}
\affiliation{Department of Physics, Institute of Science, Banaras Hindu University, Varanasi - 221005, India\\}



%% Mark off the abstract  in the ``abstract'' environment. 
\begin{abstract}
The observed large variation in the abundance of deuterium (D) in the interstellar medium (ISM) suggests that
a significant fraction of D may be depleted into polycyclic aromatic hydrocarbons (PAHs).  Signatures of the
deuteration of PAHs are expected to appear  most clearly through the C\sbond D stretching modes at 4.4--4.7\,$\mu$m,
whose strengths in emission spectra relative to those of the C\sbond H stretching modes at 3.3--3.5\,$\mu$m provide
the relative abundance of D to hydrogen (H) in PAHs, once we have accurate relative band strengths of both stretching modes.
We report experimental results of the band strengths of the C\sbond D stretching modes relative to the
C\sbond H stretching modes.  We employ a laboratory analog of interstellar carbonaceous dust, Quenched Carbonaceous Composite
(QCC), and synthesize deuterated QCC (D-QCC) by replacing the QCC starting gas of CH$_4$ with mixtures of
CH$_4$ and CD$_4$ with various ratios.  Infrared spectra of D-QCC are taken to estimate the relative
band strengths of the stretching modes, while the D/H ratios in the D-QCC samples are measured with a nanoscale
secondary ion mass spectrometer.  We obtain relative strength of aromatic and aliphatic C\sbond D to C\sbond H
stretches as $0.56 \pm 0.04$ and $0.38 \pm 0.01$ per D/H, respectively.  The ratio for the aromatic stretches is in good agreement
with the results of theoretical calculations, while that for the aliphatic stretches is smaller than for the aromatic stretches.  The present results
do not significantly change the D/H ratios in the interstellar PAHs that have previously been estimated from observed spectra.
\end{abstract}



\section{Introduction} \label{sec:intro}

Deuterium (D) is one of the elements that was created in the first minutes folllowing the Big Bang, and
its primordial abundance depends sensitively on the cosmological parameters \citep{1985ARA&A..23..319B}.
After its creation, D has been destroyed by nuclear reactions in the stellar interior, a process that is termed as astration. Its abundance 
is predicted to decrease monotonically, along with the chemical evolution of the galaxy \citep[e.g.,][]{1980ApJ...235..955M}.
Therefore, the present-day abundance of D is supposed to be directly linked to the primordial nucleosynthesis and
the subsequent chemical evolution of the galaxy.  However,  the observed ratio of D to hydrogen (H), D/H, in the galaxy shows
a large scatter, with no clear correlation with the metallicity \citep{2006ApJ...647.1106L}, which cannot be explained solely by the chemical evolution
\citep{2010IAUS..268..153T}.  \citet{2006ApJ...647.1106L} further show observational evidence that D may be depleted onto dust
grains in the interstellar medium (ISM), as originally suggested by \citet{1982NASCP2238...54J}.  
Recent observations of
distant quasars suggest that the primordial abundance of D relative to hydrogen [D/H]$_\mathrm{prim}$ is about 25\,ppm 
\citep{2018ApJ...855..102C,
2018MNRAS.477.5536Z}, while a Bayesian analysis of the observations of D in the ISM suggests that the maximum and minimum D/H ratios are
$20 \pm 1$ and $7 \pm 2$\,ppm, respectively, for a top-hat distribution \citep{2010MNRAS.406.1108P}.  These ISM values suggest that the decrease in
D abundance due to the astration is about 5\,ppm, and that D/H of about 13\,ppm is required to be depleted onto interstellar grains at maximum.  

\citet{2006ASPC..348...58D} has proposed that the major reservoir of D in the ISM could be found in polycyclic aromatic hydrocarbons (PAHs).
PAHs are thought to be the species that emit in a series of the emission bands in the near- to mid-infrared \citep{2008ARA&A..46..289T, 2020NatAs...4..339L}, which
are ubiquitously observed in the ISM \citep[e.g.,][]{1996A&A...315L.353M, 1996PASJ...48L..59O, 2013PASJ...65..120T}.
Small PAH molecules have been detected by recent radio observations with high sensitivity \citep[][and references therein]{2021NatAs...5..181B, 2021Sci...371.1265M, 
2021A&A...652L...9C}.  D enrichment is observed in interplanetary dust particles and is thought to originate in processes in interstellar clouds \citep{2000Natur.404..968M}.
\citet{2001M&PS...36.1117S} have discussed several interstellar processes for the D enrichment in PAHs \citep[see also][for recent references]{2020A&A...635A...9W}.
Assuming that the carbon (C) to total H ratio, C/H, in interstellar grains is about 200\,ppm, 
and that 85\% of carbon atoms are in the aromatic form, with H/C = 0.35 \citep{2002ApJS..138...75P},  60\,ppm of H should reside in aromatic grains
at maximum.  Note that H/C = 0.35 is an upper limit, based on small PAHs.  It is close to 0.25 for large PAHs, and the value of 60\,ppm is thus an upper limit.
Therefore, if PAHs are a major
reservoir of interstellar D, then the maximum D/H in PAHs will be about 0.2--0.3 \citep{2006ASPC..348...58D, 2014ApJ...780..114O}.

The signatures of the deuteration of PAHs should appear as shift in the wavelengths of their vibrational modes involving C\sbond D bonds \citep{1997JPCA..101.2414B}.
Aromatic and aliphatic C\sbond H stretching modes at 3.3 and 3.4--3.5\,$\mu$m shift to around 4.4 and 4.6--4.8\,$\mu$m, when deuterated, respectively \citep[e.g.,][]{2020ApJ...892...11B}.
These C\sbond D stretching modes are suggested to be the most promising features for the unambiguous detection of the deuteration, since
this spectral region is free from other PAH features \citep{2004ApJ...614..770H, 2021ApJ...917L..35A}.  \citet{1996A&A...315L.337V} report the
detection of an emission feature at 4.65\,$\mu$m with a 4.4$\sigma$ level in M17, based on observations with the Short Wavelength 
Spectrometer onboard the Infrared Space Observatory, which can be attributed to the aliphatic C\sbond D stretching vibration.
\citet{2004ApJ...604..252P} present the detection of emission bands at 4.4 and 4.65\,$\mu$m in the Orion Bar region, with 1.9$\sigma$
and 4.4$\sigma$ levels, respectively.  By comparing their intensities with the corresponding C\sbond H stretching features at 3.3--3.5\,$\mu$m, 
they estimate the D/H ratio as being $0.17 \pm 0.03$ and
$0.36 \pm 0.08$ in PAHs for the Orion Bar and M17, respectively.  These ratios are close to the value predicted by
D depletion into PAHs in the ISM.  \citet{2014ApJ...780..114O}, on the other hand, report that the ratios of the features in 4.4--4.7\,$\mu$m compared
to those in 3.3--3.5\,$\mu$m are smaller by an
order of magnitude in the Orion Bar, M17, and the reflection nebula G18.14.0, based on observations with AKARI.
They estimate D/H as being 3\% at most in interstellar PAHs, assuming a factor of a 0.57 difference in the cross section between fully deuterated (perdeuterated) and
fully hydrogenated PAHs, based on Density Functional Theory
(DFT) calculations \citep{1997JPCA..101.2414B} and different excitation conditions between the 3 and 4\,$\mu$m bands.  
\citet{2016A&A...586A..65D} further search for features between 4.4 and 4.8\,$\mu$m in AKARI spectra of \ion{H}{2} regions,
finding the features in only six Galactic sources out of 41.

Since the PAHs in the ISM are supposed to be only partially deuterated, the band strengths of the C\sbond D stretching modes of PAHs with various D/H ratios
are required to correctly estimate D/H from observed spectra.  DFT calculations of singly deuterated and deuteronated PAHs of various kinds have been 
calculated, which indicate that the band strength ratio of the C\sbond D to C\sbond H stretches per unit bond varies with the PAH size \citep{2015MNRAS.454..193B,
2016P&SS..133...97B, 2020ApJ...892...11B}.  \citet{2020ApJS..251...12Y, 2021ApJS..255...23Y} estimate the average band strength ratio of aromatic
C\sbond D to C\sbond H stretching modes as $\sim 0.56$, based on
DFT calculations of a number of monodeuterated and multideuterated neutral PAHs of relatively small size. 
With the average value, they estimate that the degree of deuteration of PAHs (the fraction of peripheral atoms
attached to carbon atoms in the form of D) is $\sim 2.4$\%.

Laboratory measurements of infrared spectra have only been made for perdeuterated PAHs \citep{doi:10.1021/j100067a008, 1997JPCA..101.2414B, 2020ApJS..251...12Y}.
In this paper, we experimentally investigate the effect on the band intensities of partial deuteration by using a laboratory organic dust analog, Quenched Carbonaceous Composite (QCC).
QCC is an organic material synthesized from methane (CH$_4$) plasma, which shows the infrared spectrum that reproduces the observed emission features from 3 to 14\,$\mu$m fairly well
\citep{1984ApJ...287L..51S, 1987ApJ...320L..63S, 1990ApJ...353..543S}.  By replacing the starting gas of CH$_4$ with mixtures of CH$_4$ and CD$_4$ with various ratios, we
synthesize partially deuterated QCCs with different D/H contents (hereafter called D-QCCs).   The actual content of D relative to H in a D-QCC is measured by a nanoscale secondary ion mass spectrometer (NanoSIMS),
and the intensity ratio of the bands in 4.4--4.7\,$\mu$m to 3.3--3.5\,$\mu$m is compared with the D/H ratio in the sample.

The experiment's method and results are described in \S~\ref{sec:exp}, and the results are discussed in \S~\ref{sec:dis}.  
A summary is given in \S~\ref{sec:sum}.

\section{Experiment method and results}\label{sec:exp}

The experimental method for producing D-QCC is the same as that for synthesizing QCC \citep{1983Natur.301..493S} except that the starting gas of CH$_4$ is replaced by mixtures of
CH$_4$ and CD$_4$ with various ratios.  A similar method has been employed to study the effect of the $^{13}$C isotope in QCC \citep{2003A&A...407..551W}.
The starting gas is introduced into a vacuum chamber and applied by a microwave discharge, to produce plasma.  The plasma
gas is quenched in the quartz tube, and the product (D-QCC) is collected on KBr and Si substrates in the tube.
We use the starting gases with CD$_4$ fractions of 0.0 (fully CH$_4$ gas), 19.7, 38.8, 79.4, and 100.0\% (fully CD$_4$ gas).
We designate the products QCC, D-QCC20, D-QCC40, D-QCC80, and D-QCC100, respectively, in the following.  

\subsection{Infrared band measurement}

\begin{figure}[ht]
\epsscale{0.6}
\plotone{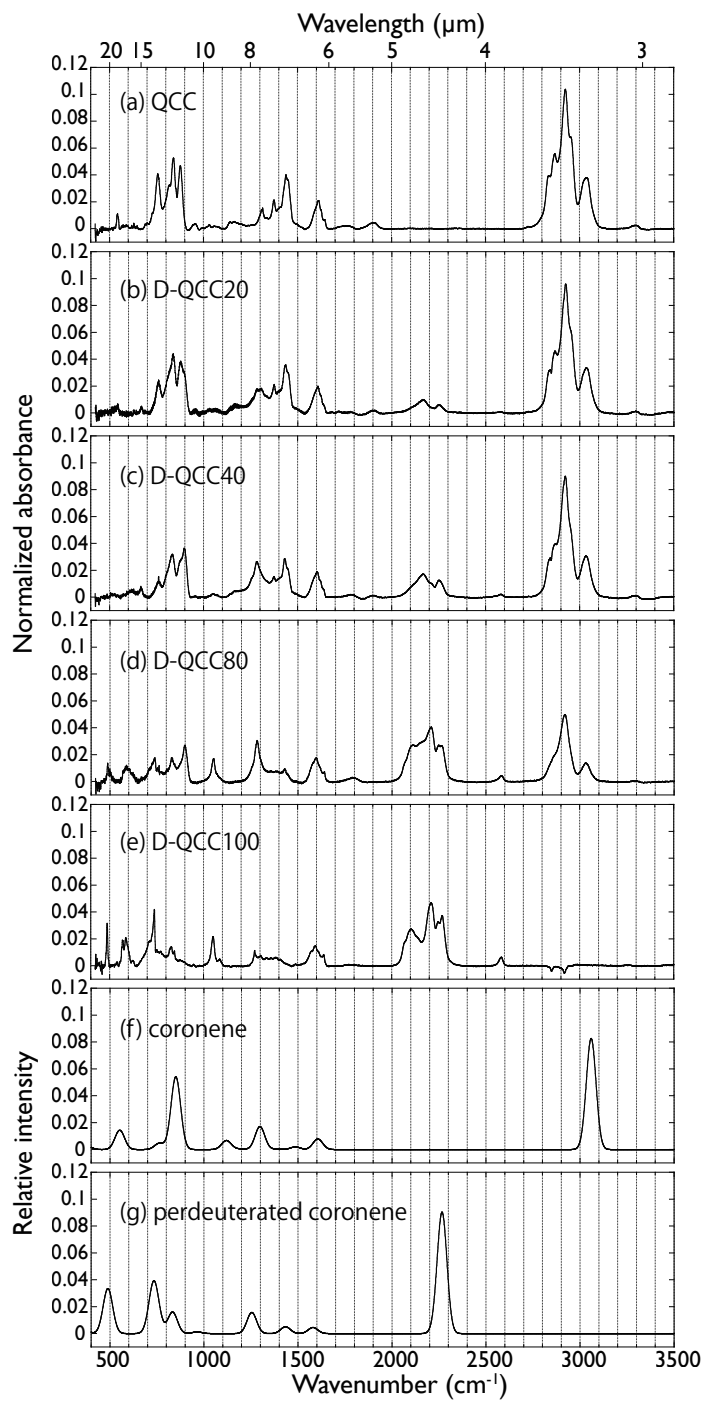}
\caption{Absorbance spectra of QCC and D-QCC samples and the results of DFT calculations.  The continuum components fitted by spline functions are subtracted from 
the spectra.  They are further normalized by the integrated strength of the feature at 1610\,cm$^{-1}$ (6.2\,$\mu$m) for 
illustrative purposes. Panels (a), (b), (c), (d), and (e) show the spectra of QCC, D-QCC20, D-QCC40, D-QCC80, and D-QCC100, respectively. 
Panels (f) and (g) show the results of the DFT calculations for coronene (C$_{24}$H$_{12}$) and perdeuterated coronene (C$_{24}$D$_{12}$), respectively, as reference for the identification of the features associated with D.
The features at 2770--3110\,cm${-1}$ (3.2--3.6\,$\mu$m) in QCC (a) are attributed to C\sbond H stretching modes, while the new features at 2000--2350\,cm$^{-1}$ (4.25--5.0\,$\mu$m), which are attributed to C\sbond D stretching modes,
become stronger with increasing D/H (b--e).  The weak features at around 500 and 700\,cm$^{-1}$ (20 and 14\,$\mu$m) are 
seen in the high-D/H cases (d and e), and corresponding features are also seen in perdeuterated coronene (g).  The wavelength scale is given 
in the upper panel.
\label{fig:IRsp}}
\end{figure}

Infrared spectra of QCC and D-QCC were measured using the sample collected on the KBr substrate  for the range 400--3500\,cm$^{-1}$ (2.9--25\,$\mu$m), with a resolution of 2\,cm$^{-1}$
by the Fourier Transform Infrared Spectroscopy (FT-IR) spectrometer IFS 125 of Bruker Optics.  The infrared spectrum of a blank KBr substrate was taken for each measurement,
to be used as a reference.
All the measurements were carried out over an area with a diameter of 5\,mm of the substrate.  The transmission spectra were converted into absorbance
spectra, after being divided by the reference spectra.  A spline function was fit to the continuum component of each absorbance spectrum, and subtracted to extract the band features.
The thickness of D-QCC was not the same for all the samples.  The absorbance spectra are normalized by 
the integrated band strength of the feature at 1610\,cm$^{-1}$ (6.2\,$\mu$m), which is attributed to a C\sbond C vibration
and is not supposed to be affected directly by deuteration.  The normalization is made only for the illustrative purposes, to
indicate the relative variation of each feature.  The resultant spectra are shown in Figure~\ref{fig:IRsp}.  The results of the DFT calculations for
coronene (C$_{24}$H$_{12}$) and perdeuterated coronene (C$_{24}$D$_{12}$) \citep[e.g.,][]{2015MNRAS.454..193B} are also shown, as reference for the identification of the features associated with D.
The DFT spectra are smoothed with a FWHM of 60\,cm$^{-1}$ for the illustrative purposes.

Figure~\ref{fig:IRsp} shows that the C\sbond H stretching features at 2770--3110\,cm$^{-1}$ (3.2--3.6\,$\mu$m) gradually decrease with the increase of the
fraction of D in D-QCC, while new features appear at 2000--2350\,cm$^{-1}$ (4.25--5.0\,$\mu$m), which are attributed to C\sbond D stretching modes,
indicated by the spectra of coronene and perdeuterated coronene.  The features in the range 1200--1700\,cm$^{-1}$ (5.9--8.3\,$\mu$m) are relatively unchanged
by deuteration, as expected, although their relative intensities vary with the fraction of D.  For the frequency range lower than 1000\,cm$^{-1}$ ($ > 10$\,$\mu$m), the appearance
of the features becomes complex, as suggested by theoretical calculations \citep[e.g.,][]{2004ApJ...614..770H, 2021ApJ...917L..35A}, making it 
difficult to extract reliable information regarding deuterated PAHs.  There are a few features that are only seen in D-QCC80
and D-QCC100, which have corresponding features in perdeuterated coronene, e.g., at around 500 and 700\,cm$^{-1}$ (20 and 14\,$\mu$m), respectively, but they appear to be faint in D-QCCs with lower D/H.

\begin{figure}[ht!]
\epsscale{0.75}
\plotone{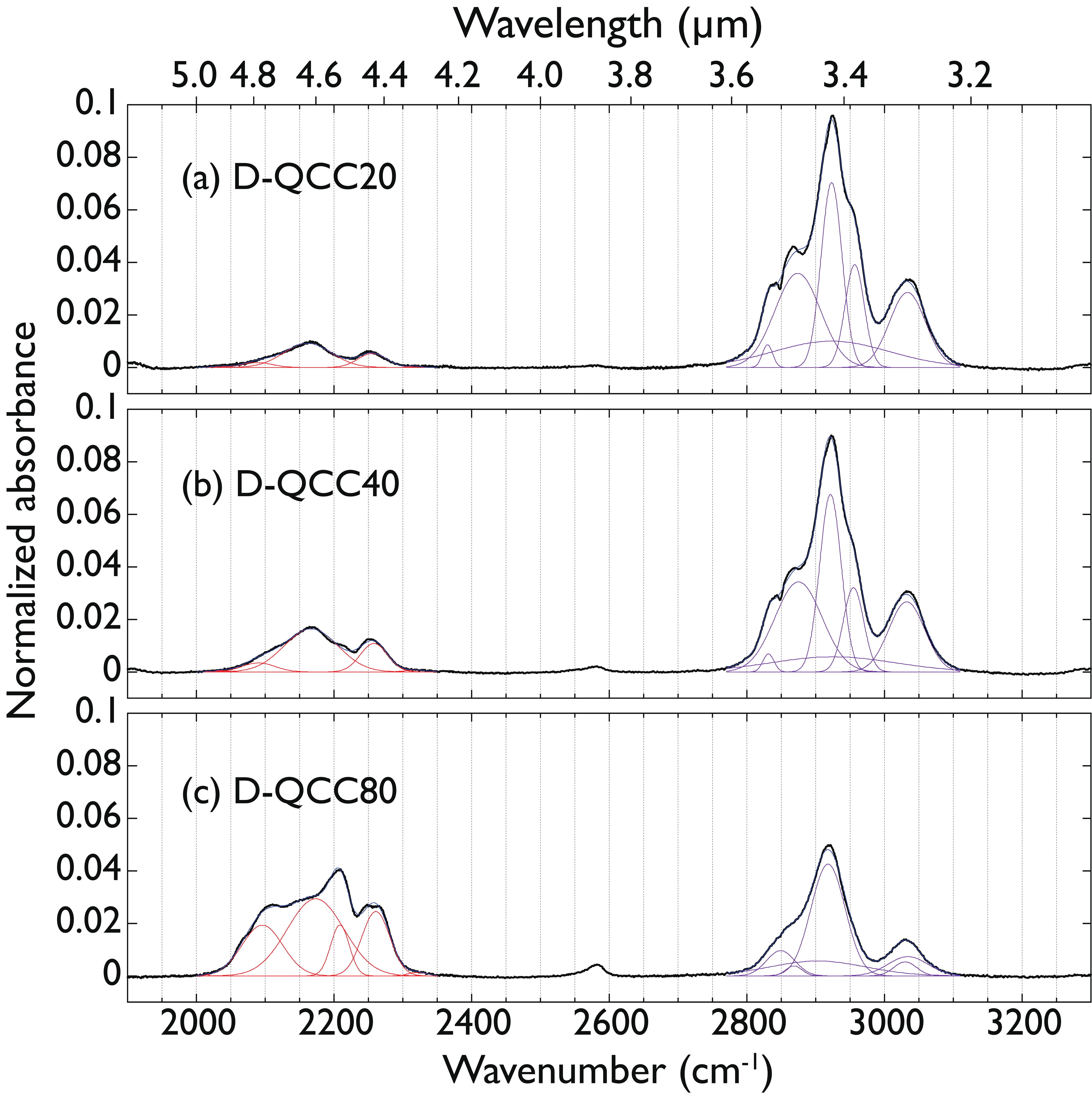}
\caption{Enlarged figures of normalized absorbance for the 1900--3300\,cm$^{-1}$ (3.0--5.3\,$\mu$m) regions of (a) D-QCC20, (b) D-QCC40, and (c) D-QCC80.  
The features between 2000--2350\,cm$^{-1}$ (4.25--5.0\,$\mu$m) and 2770--3110\,cm$^{-1}$ (3.2--3.6\,$\mu$m) are decomposed into five to six Gaussians, which are shown by the thin red and
purple thin lines, respectively.  The summation of the Gaussians is shown by the blue lines, which overlap with the measured absorbance of the black lines. The wavelength scale is given 
in the upper panel.
\label{fig:decomp}}
\end{figure}

While the features around 3.3 and 4.4\,$\mu$m are attributed to aromatic C\sbond H and C\sbond D stretching modes,
those at 3.4--3.5 and 4.6--4.8\,$\mu$m are ascribed to the stretching modes of aliphatic side-groups of C\sbond H and C\sbond D, respectively
\cite[e.g.,][]{2020ApJ...892...11B}.  Since the band ratios of the C\sbond H and C\sbond D stretches could be different between aromatic
and aliphatic stretches, we analyze the aromatic and aliphatic stretches separately.  The features in 2000--2350\,cm$^{-1}$  (4.25--5.0\,$\mu$m) 
and 2770--3110\,cm$^{-1}$ (3.2--3.6\,$\mu$m) in
the absorbance spectra are decomposed into five to six Gaussians, as shown in Figure~\ref{fig:decomp}.  The Gaussian components whose
central frequencies are located between 2000--2250\,cm$^{-1}$ (4.4--5.0\,$\mu$m) and 2770--3000\,cm$^{-1}$ (3.3--3.6\,$\mu$m) are assigned to aliphatic 
bonds of C\sbond D and C\sbond H stretching modes, while those with the central frequencies between 2250--2350\,cm$^{-1}$ (4.25--4.4\,$\mu$m)
and 3000--3110\,cm$^{-1}$ (3.2--3.3\,$\mu$m)
are attributed to aromatic bonds of the C\sbond D and C\sbond H stretching modes, respectively.  Note that only one Gaussian is used to fit the aromatic stretches for both the
aromatic C\sbond H and C\sbond D bonds.  Figure~\ref{fig:decomp} shows that the decomposition is good (errors $ < 2$\%)
and that the separation of aliphatic from aromatic bonds is made unambiguously.  We sum up the integrated strength of each Gaussian
and estimate the band strength ratios of the C\sbond D to C\sbond H stretching modes for the aromatic and aliphatic bonds separately.
The C\sbond D to C\sbond H band strengths ratios are separately summarized for aromatic and aliphatic components in Table~\ref{tab:band}.
The spectrum of D-QCC80 appears to be notably different from those of D-QCC20 and D-QCC40 for wavenumbers of 2000--2250\,cm$^{-1}$, whose
range corresponds to aliphatic C\sbond D stretching vibrations.  We surmise that the difference originates from an increase in the combinations of symmetric and asymmetric 
aliphatic C\sbond D stretching vibrations,
since a large fraction of H in the aliphatic bonds is replaced by D in D-QCC80.

\subsection{D/H measurement}

The D/H ratio of the sample on the Si substrate was measured with the Cameca NanoSIMS 50 ion microprobe at the Atmosphere and Ocean Research Institute, 
University of Tokyo.  The radius and thickness of the Si substrate were 5\,mm and 0.7\,mm, respectively.  
We employed a coronene (C$_{24}$H$_{12}$) sample that had a known D/H ratio ($\delta$D relative to the Vienna Standard Mean Ocean Water (VSMOW) 
equal to $-45.7$\%) for the calibration.  The sample and the coronene were measured under the same experimental conditions.   
We also measured a blank Si substrate to
determine the blank level.  The blank hydrogen level was less than 1\% of the sample level.
The sample was presputtered for a 20 $\times$ 20\,$\mu$m area, with a Cs$^+$ beam of 200\,pA, to remove contamination on the surface.  
Then, the measurements of H$^-$ and D$^-$ were carried out for the central area of
a 5 $\times$ 5\,$\mu$m with a beam of 2.5\,pA, for 200\,s ,to obtain the D/H ratio.
Since the size of the NanoSIMS measurement area was small compared to the infrared spectroscopy, 
we selected positions that produced sufficient signals over a region of a similar size  to the infrared measurements, then took the average.  
Four positions distributed over a $\sim 5$\,mm diameter region were measured
for each of the D-QCC samples.  The variation of D/H was less than 10\% among the measured positions,
suggesting that D/H is fairly uniform over the sample and that the average value represents the D/H ratio of the sample reliably.  
The results are also summarized in Table~\ref{tab:band}.   D-QCC20 has almost the same D/H ratio as that of the starting gas, while
the D/H ratios of D-QCC40 and D-QCC80 are smaller than that of the starting gas, suggesting that D is incorporated into D-QCC less efficiently than H
in the present synthesis process.

\begin{deluxetable*}{ccccc}[h!]
\tablecaption{C\sbond D to C\sbond H Band Strength Ratios and D/H Ratios of D-QCCs\label{tab:band}}
\tablewidth{0pt}
\tablehead{
\colhead{Sample} &  \multicolumn2c{C\sbond D to C\sbond H Band Strength Ratio} & \colhead{Gas D/H} & \colhead{D-QCC D/H} \\
\colhead{} & \colhead{Aliphatic Stretch}& \colhead{Aromatic Stretch}
& \colhead{} &  \colhead{} }
\startdata
D-QCC20 & $0.098 \pm 0.001$ & $0.154\pm 0.001$ & 0.245 & $0.24 \pm 0.02$ \\
D-QCC40 & $0.222 \pm 0.002$ & $0.323 \pm 0.003$ & 0.634 & $0.60 \pm 0.03$\\
D-QCC80 & $1.21 \pm 0.03$ & $1.72 \pm 0.05$ & 3.85 & $3.2 \pm 0.2$\\
\enddata
\end{deluxetable*}

\section{Discussion}\label{sec:dis}

\begin{figure}[ht!]
\epsscale{0.7}
\plotone{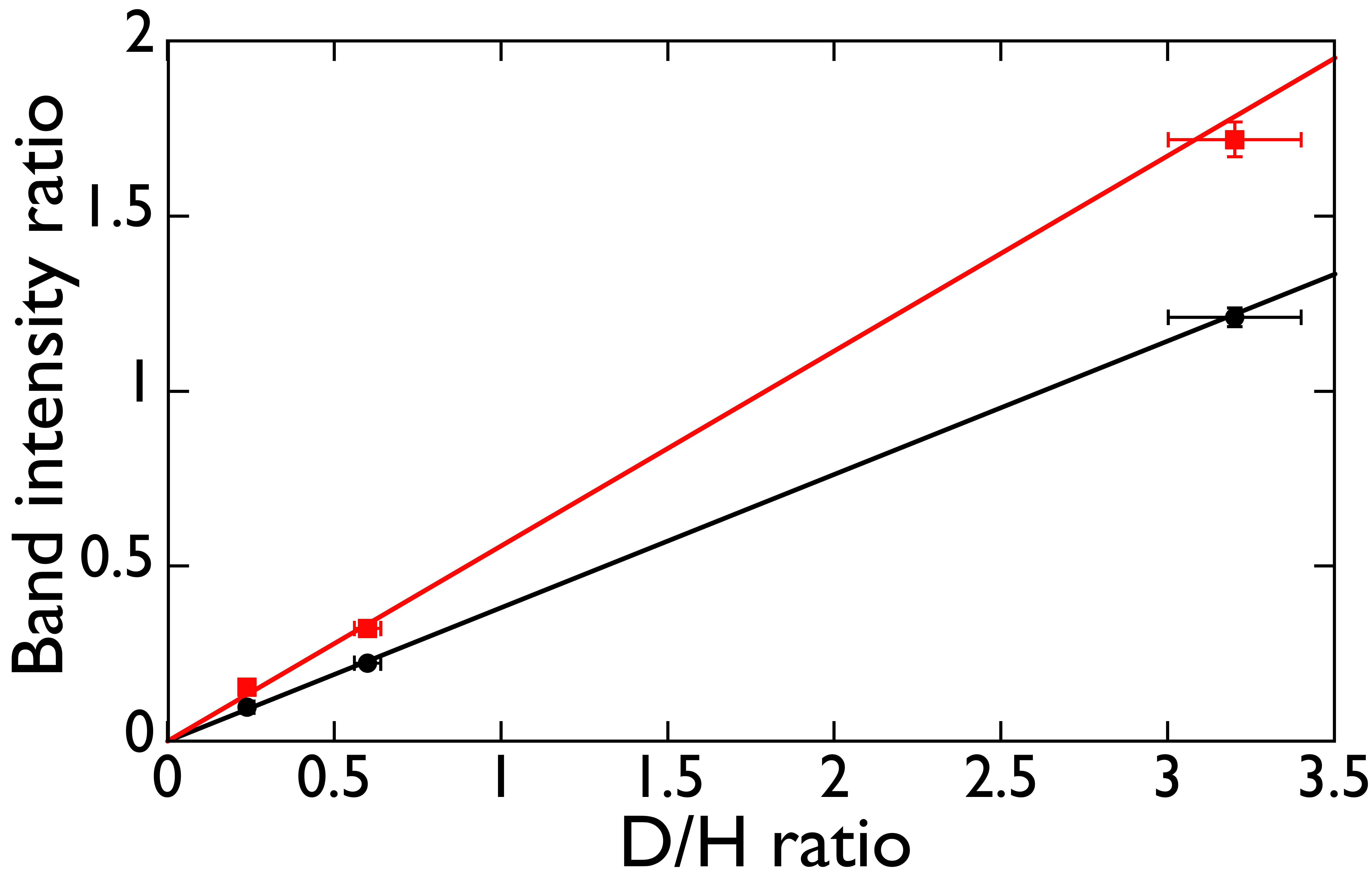}
\caption{Band strength ratios of C\sbond D to C\sbond H for the aliphatic and aromatic stretching modes against the D/H ratio of the sample measured by the
NanoSIMS.
The red squares indicate the data for the aromatic bonds, while the black circles show those for the aliphatic bonds.  The red and black straight lines
show the best-fit lines for the aromatic and aliphatic bonds, respectively.
\label{fig:band}}
\end{figure}

The band strength ratios of the C\sbond D to C\sbond H stretching modes are plotted against the D/H ratio of the sample in Figure~\ref{fig:band}.  The data are well fitted with straight lines, 
which show the band strength per D/H as being $0.56 \pm 0.04$ and $0.38 \pm 0.01$ for the aromatic and aliphatic bonds, respectively.  The value for
the aromatic bonds is in good agreement with the theoretical value for perdeuterated PAHs of 0.57 \citep{1997JPCA..101.2414B} and the average ratios for
monodeuterated and multideuterated neutral PAHs of $\sim 0.56$ \citep{2020ApJS..251...12Y, 2021ApJS..255...23Y}. Therefore, the present
experimental results will not change the conclusions of
previous studies on the D fraction in the aromatic bonds.  On the other hand, the present study indicates that
the band strength ratios of the aliphatic bonds are about two-thirds of the ratios of the aromatic bonds.  While \citet{2020ApJ...892...11B}
discussed aliphatic C\sbond D bands of several PAHs based on DFT calculations, no theoretical studies have so far been made for the band strength ratios 
of the aliphatic bonds.  In this analysis, we simply assume
that D is distributed among the aliphatic and aromatic bonds in the same fraction, i.e., the D/H ratio is the same for both bonds.
The agreement of the experimental results for the aromatic bonds suggests that this assumption is not largely in error.

Astronomical observations detect larger excess emission attributable to the aliphatic C\sbond D stretching modes at $\sim 4.6$\,$\mu$m compared to the aromatic C\sbond D modes at $\sim 4.4$\,$\mu$m
\citep{2004ApJ...604..252P, 2014ApJ...780..114O, 2016A&A...586A..65D}, suggesting that D may reside more in aliphatic bonds than in aromatic ones.
\citet{2004ApJ...604..252P} and \citet{2014ApJ...780..114O} assume the same band ratio for both the aromatic and aliphatic bonds, and estimate the D/H ratio
from the summation of the aromatic and aliphatic band intensities.  In the following, we roughly estimate the D content in the aliphatic bonds,
based on the AKARI observations of the Orion Bar, where the ratio of aliphatic C\sbond D to C\sbond H stretching 
features appears to be the largest  among the three targets \citep{2014ApJ...780..114O}.

\citet{2014ApJ...780..114O} report that the band intensity ratios of the C\sbond D to C\sbond H stretching modes for the aromatic and aliphatic bonds are $0.026 \pm 0.002$ and
$0.04 \pm 0.007$, respectively, in the Orion Bar.  The emission of the 3.3\,$\mu$m band requires higher excitation than that of the 4.4\,$\mu$m, 
the difference inf which depends on
the size of the PAHs and the spectrum of the incident radiation field.
Assuming that the relative band strength for the aromatic C\sbond D to C\sbond H stretches is 0.56, and
that the difference in the excitation is about a factor of 1.75 between 4.4 and 3.3\,$\mu$m, as in \citet{2014ApJ...780..114O}, we then obtain D/H in the aromatic bonds as $0.026 \pm 0.002$.  A similar
estimate for the aliphatic bonds results in D/H of $0.058 \pm 0.010$, assuming a relative band strength of 0.38 for the aliphatic bonds, as obtained in the present study, and the same factor
of the difference in the excitation efficiency as for the aromatic bonds.  The number of C atoms in aliphatic units is estimated to be smaller than the number in aromatic units \citep[e.g.,][]{2016ApJ...825...22Y, 2017NewAR..77....1Y}.  
Assuming an average ratio of the band strength of the aliphatic to aromatic C\sbond H stretches of 1.76, estimated from various aliphatic
side-groups \citep{2016ApJ...825...22Y}, and the observed band ratio of the Orion Bar (0.36), we estimate that the H fraction in the aliphatic bonds in the Orion
Bar is $0.205 \pm 0.004$.  The total D/H ratio of the aromatic and aliphatic bonds becomes $0.031 \pm 0.005$, which is within the uncertainty of the D/H ratio  of
$0.029 \pm 0.002$, estimated without separating the aromatic and aliphatic bonds.  
Therefore, the band strength ratio obtained in the present study does not change the D/H ratio estimated in previous studies.
Note that these estimates do not include the uncertainties in the band strength ratios.  

The present experiments were made with relatively high D/H ratios compared to those estimated from observations.
The features of the C\sbond D vibration modes become weak, and it is difficult to make an accurate experimental estimate for low-D/H samples.
The good linearity seen in Figure~\ref{fig:band} and the agreement of the numerical results with the DFT calculations
for the band strength ratio of the aromatic bonds, suggest that the band strength ratios derived in the present experiments should also be valid even for
low-D/H cases, and are applicable to interpretations of astronomical observations.  

\section{Summary}\label{sec:sum}
In the present study, we experimentally investigate the relative band strength ratios of the C\sbond D to C\sbond H
stretching modes.  By replacing the starting gas of CH$_4$ with mixtures of CH$_4$ and CD$_4$ gas in the synthesis of QCC, which reproduces most of
the observed emission features from 3 to 14\,$\mu$m \citep{1984ApJ...287L..51S}, we successfully produce partially
deuterated QCC (D-QCC) with various D/H ratios.  The D/H ratios of D-QCC are measured with the NanoSIMS.
We estimate the band strength ratios for the aromatic and aliphatic bonds separately, both of which show a good linearlity with
the D/H ratio.
The band strength ratio for the aromatic bonds per D/H is found to be $0.56 \pm 0.04$, which is in good agreement with numerical calculations
of monodeuterated and multideuterated neutral PAHs of small sizes of 0.56--0.57 \citep{2020ApJS..251...12Y,
2021ApJS..255...23Y}.  The band ratio for the aliphatic bonds, on the other hand, shows a smaller ratio of $0.38 \pm 0.01$.
Since observed spectra suggest the dominance of aromatic bonds in the band carriers, the present results will not affect
the D/H ratio in PAHs that has been previously estimated from observations.
The present study confirms the conclusions of previous studies, that the D/H ratios in PAHs that emit at 3.3--3.5\,$\mu$m
are small and that the missing D in the ISM cannot be explained by depletion into these PAHs  \citep{2014ApJ...780..114O, 2021ApJS..255...23Y}.  
The degree of deuteration of the PAHs depends on the environment {\citep{2021ApJ...917L..35A}}.  Most of the observed spectra of the 3.3--4.7\,$\mu$m
emission features are taken toward photodissociation regions, where the gas temperature is relatively high.  The missing D may be depleted into PAHs
in colder regions, without a sufficient amount of ultraviolet photons, where the emission of 3.3--4.7\,$\mu$m is not excited efficiently.  The features of deuterated PAHs at longer wavelengths may
have to be investigated.  However, the present study also confirms that the features of deuterated PAHs at longer wavelengths are confused with
other PAH features, as predicted, and suggests that their unambiguous detection requires highly sensitive spectroscopy, because the features unique to
D at 14--20\,$\mu$m appear to be faint for low-D/H ratios.

\acknowledgments
%The authors thank N. Takahata and Y. Sano at Atmosphere and Ocean Research Institute, the University of Tokyo, for
%allowing us to use their NanoSIMS 50 ion microprobe and their help in the nanoSIMS measurements.
The authors thank Setsuko Wada and Seiji Kimura at the University of Electro-Communications for their help with
the QCC experiments.
This work was supported by JSPS KAKENHI grant Nos. JP16H00934 and JP18K03691.
T.M. and M.B. acknowledge financial support from the JSPS fellowship.
A.P. acknowledges financial support from an IoE grant of Banaras Hindu University (R/Dev/D/IoE/Incentive/2021-22/32439) and
 financial support through the Core Research Grant of SERB, New Delhi (CRG/2021/000907). 
The authors also acknowledge support from a DST--JSPS grant (DST/INT/JSPS/P-238/2017).

\bibliography{PAD-exp}{}
\bibliographystyle{aasjournal}



\end{document}